\documentclass{elsart3p}

\usepackage{graphicx}

\usepackage{amssymb}
\usepackage{amsmath}
\usepackage{bm}

\begin{document}

\begin{frontmatter}



\title{Spin-resolved crossed Andreev reflection in ballistic heterostructures}


\author[int,otf]{Mikhail S. Kalenkov \thanksref{thank1}},
\author[int,otf]{Andrei D. Zaikin}

\address[int]{Forschungszentrum Karlsruhe, Institut f\"ur Nanotechnologie,
76021, Karlsruhe, Germany}
\address[otf]{I.E. Tamm Department of Theoretical Physics, P.N.
Lebedev Physics Institute, 119991 Moscow, Russia}
\thanks[thank1]{Corresponding author.  E-mail: kalenkov@lpi.ru}

\begin{abstract}
We theoretically analyze non-local effects in electron transport
across three-terminal ballistic normal-superconducting-normal (NSN)
structures with spin-active interfaces. Subgap electrons entering
S-electrode from one N-metal may form Cooper pairs with their
counterparts penetrating from another N-metal. This phenomenon of
crossed Andreev reflection is highly sensitive to electron spins
and yields a rich variety of properties of  non-local conductance
which we describe non-perturbatively at arbitrary interface
transmissions, voltages and temperatures. Our results can be
applied to hybrid structures with normal, ferromagnetic and
half-metallic electrodes and can be directly tested in future
experiments.
\end{abstract}

\end{frontmatter}


\section{Introduction}

Andreev reflection (AR) \cite{And} is the main mechanism of low
energy electron transport between a normal metal and a
superconductor. This mechanism results in a number of interesting
effects causing, e.g., a non-zero subgap conductance \cite{BTK} of
such hybrid structures. In systems with one superconducting (S)
and several normal (N) terminals, e.g. NSN hybrid structures,
electrons may suffer Andreev reflection at each of NS interfaces.
Provided the distance between two NS interfaces $L$ strongly
exceeds the superconducting coherence length $\xi$, AR processes
at these interfaces are independent. If, however, the distance $L$
becomes comparable with $\xi$, two additional {\it non-local}
processes should be taken into account (see Fig. 1). Firstly, an
electron with subgap energy can directly penetrate from one
N-metal into another through a superconductor. Since the subgap
density of states in the superconductor vanishes, the probability
of this process should be suppressed by the factor $\sim \exp
(-L/\xi )$. Secondly, an electron penetrating into the
superconductor from the first N-terminal may form a Cooper pair
together with another electron from the second N-terminal. In this
case a hole goes into the second N-metal and AR becomes a
non-local effect. The probability of this process -- usually
called crossed Andreev reflection (CAR) \cite{BF,GF} -- also
decays as $\sim \exp (-L/\xi )$ and, in combination with direct
electron transfer between normal electrodes, determines non-local
conductance in hybrid multi-terminal structures. This non-local
conductance can be directly measured in experiments and it has
recently become a subject of intensive investigations.

\begin{figure}
\includegraphics[width=75mm]{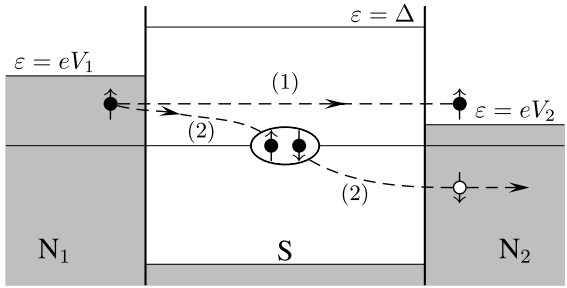}
\caption{Two elementary processes contributing to non-local
conductance of an NSN device: (1) direct electron transfer and (2)
crossed Andreev reflection.}
\end{figure}

Several experiments \cite{Beckmann,Teun,Venkat} allowed to clearly
detect the non-local conductance in three-terminal NSN structures
and demonstrated a rich variety of different features which
unambiguous and detailed interpretation remains an important task
for the future. At this point we note that in addition to CAR a
number of other physical effects may considerably influence the
observations. Among such effects we mention, e.g, charge imbalance
(relevant close to the superconducting critical temperature
\cite{Beckmann,Venkat}) and zero-bias anomalies in the Andreev
conductance due to both disorder-enhanced interference of
electrons \cite{VZK,HN,Z} and Coulomb effects \cite{Z,HHK,GZ06}.

Theoretically CAR was analyzed within the perturbation theory in
the transmission of NS interfaces in Refs. \cite{FFH,Fabio} where
it was demonstrated that in the lowest order in the interface
barrier transmission and at $T=0$ CAR contribution to
cross-terminal conductance is exactly canceled by that from
elastic electron cotunneling (indicated as (1) in Fig. 1), i.e.
the non-local conductance turns out to vanish in this limit.
Recently a non-trivial interplay between normal reflection,
tunneling, local AR and CAR in three-terminal ballistic NSN
devices was non-perturbatively analyzed to all orders in the
interface transmissions \cite{KZ06}. This analysis allowed to
determine an explicit dependence of the non-local conductance both
on the transmissions of NS interfaces and on the length $L$, to
set the maximum scale of the effect and to consider various
important limits. The effect of disorder on CAR was recently
studied in Refs. \cite{BG} (perturbatively in tunneling) and
\cite{Belzig} (non-perturbatively in tunneling, for a device with
normal terminals attached to a superconductor via an additional
normal island). The interplay between CAR and Coulomb interaction
effects was recently addressed in Refs.  \cite{LY,GZ07}.

It is also important to mention that both AR and CAR should be
sensitive to magnetic properties of normal electrodes because
these processes essentially depend on spins of
scattered electrons. First experiments on
ferromagnet-superconductor-ferromagnet (FSF) structures
\cite{Beckmann} illustrated this point by demonstrating the
dependence of non-local conductance on the polarization of
ferromagnetic terminals. Hence, for better understanding of
non-local effects in multiterminal hybrid proximity structures it
is desirable to construct a theory of {\it spin-resolved} CAR. In
the lowest order order in tunneling this task was accomplished in
Ref. \cite{FFH}. For FSF structures higher orders in the interface
transmissions were considered in Refs. \cite{MF,Melin}.

In this paper we are going to generalize our quasiclassical approach
\cite{KZ06} and construct a theory of spin-resolved CAR to all orders
in the interface transmissions. Instead of dealing directly with FSF
devices we will consider NSN structures with spin-active interfaces.
This model allows to distinguish spin-dependent contributions to the
non-local conductance and to effectively mimic the situation of
ferromagnetic and/or half-metallic electrodes.

The structure of the paper is as follows. In Sec. 2 we will introduce our
model and discuss the quasiclassical formalism supplemented by the
boundary conditions for Green-Keldysh functions which account for
electron scattering at spin-active interfaces. Non-local electron
transport in NSN structures with such interfaces will be analyzed in Sec. 3.
Our main conclusions will be briefly summarized in Sec. 4. Technical details
related to boundary conditions will be outlined in Appendix A.

\begin{figure}
\centerline{\includegraphics[width=65mm]{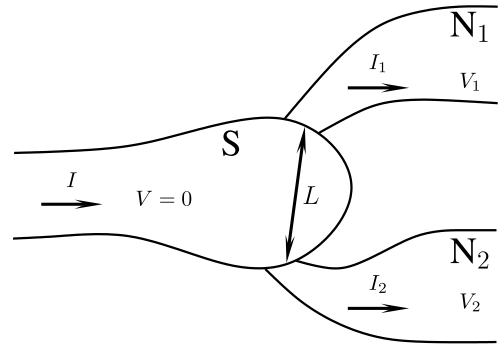}}
\caption{Schematics of our NSN device.} \label{nsn}
\end{figure}

\section{The model and formalism}
Let us consider three-terminal NSN structure depicted in Fig. \ref{nsn}.
We will assume that all three metallic electrodes are non-magnetic
and ballistic, i.e. the electron elastic mean free path in each
metal is larger than any other relevant size scale. In order
to resolve spin-dependent effects we will assume that both
NS interfaces are spin-active, i.e. we will distinguish ``spin-up''
and ``spin-down'' transmissions of the first ($D_{1\uparrow}$ and
$D_{1\downarrow}$) and the second  ($D_{2\uparrow}$ and
$D_{2\downarrow}$) SN interface. All these four transmissions
may take any value from zero to one. The effective cross-sections of
the two interfaces will be denoted respectively as
$\mathcal{A}_1$ and $\mathcal{A}_2$. The distance between these
interfaces $L$ as well as
other geometric parameters are assumed to be much larger than
$\sqrt{\mathcal{A}_{1,2}}$, i.e. effectively both contacts are
metallic constrictions. In this case the voltage drops only across
SN interfaces and not inside large metallic electrodes. Hence,
nonequilibrium (e.g. charge imbalance) effects related to the
electric field penetration into the S-electrode can be neglected.
In what follows we will also ignore Coulomb effects
\cite{Z,HHK,GZ06}.

For convenience, we will set the electric potential of the
S-electrode equal to zero, $V=0$. In the presence of bias voltages
$V_1$ and $V_2$ applied to two normal electrodes (see Fig.
\ref{nsn}) the currents $I_1$ and $I_2$ will flow through SN$_1$
and SN$_2$ interfaces. These currents can be evaluated with the
aid of the quasiclassical formalism of nonequilibrium
Green-Eilenberger-Keldysh functions $\hat g^{R,A,K}$ \cite{BWBSZ}
which we briefly specify below.

\subsection{Quasiclassical equations}

In the ballistic limit the corresponding equations take the form
\begin{gather}
\begin{split}
\left[
\varepsilon \hat\tau_3+
eV(\bm{r},t)-
\hat\Delta(\bm{r},t),
\hat g^{R,A,K} (\bm{p}_F, \varepsilon, \bm{r},t)
\right]
+\\+
i\bm{v}_F \nabla \hat g^{R,A,K} (\bm{p}_F, \varepsilon, \bm{r},t) =0,
\end{split}
\label{Eil}
\end{gather}
where $[\hat a, \hat b]= \hat a\hat b - \hat b \hat
a$, $\varepsilon$ is the quasiparticle energy, $\bm{p}_F=m\bm{v}_F$ is the
electron Fermi momentum vector and $\hat\tau_3$ is the Pauli matrix in Nambu
space.
The functions  $\hat g^{R,A,K}$ also obey the normalization conditions
$(\hat g^R)^2=(\hat g^A)^2=1$ and $\hat g^R \hat g^K + \hat g^K \hat g^A =0$.
Here and below the product of matrices is defined as time convolution.

Green functions $\hat g^{R,A,K}$ and $\hat\Delta$ are $4\times4$ matrices in
Nambu and spin spaces. In Nambu space they can be parameterized as
\begin{equation}
        \hat g^{R,A,K} =
        \begin{pmatrix}
                g^{R,A,K} & f^{R,A,K} \\
                \tilde f^{R,A,K} & \tilde g^{R,A,K} \\
        \end{pmatrix}, \quad
        \hat\Delta=
        \begin{pmatrix}
                0 & \Delta i\sigma_2 \\
                \Delta^* i\sigma_2& 0 \\
        \end{pmatrix},
\end{equation}
where $g^{R,A,K}$, $f^{R,A,K}$, $\tilde f^{R,A,K}$, $\tilde g^{R,A,K}$ are
$2\times2$ matrices in the spin space, $\Delta$ is the BCS order parameter and
$\sigma_i$ are Pauli matrices. For simplicity we will only consider the case of
spin-singlet isotropic pairing in the superconducting electrode.
The current density is
related to the Keldysh function $\hat g^K$ according to the standard formula
\begin{equation}
\bm{j}(\bm{r}, t)= -\dfrac{e N_0}{8} \int d \varepsilon
\left< \bm{v}_F \mathrm{Sp} [\hat \tau_3 \hat g^K(\bm{p}_F,
\varepsilon, \bm{r},t)] \right>,
\label{current}
\end{equation}
where $N_0=mp_F/2\pi^2$ is the density of state at the Fermi level and
angular brackets $\left< ... \right>$ denote averaging over the Fermi momentum.

\subsection{Riccati parameterization}

\begin{figure*}
\centerline{
\includegraphics{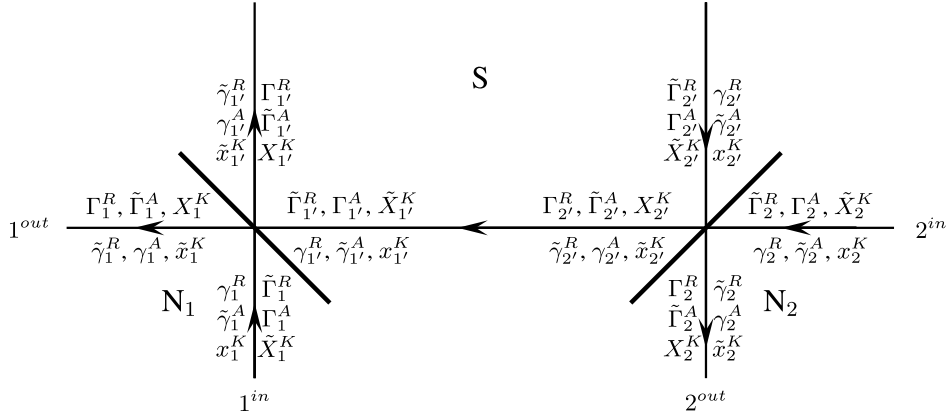}
} \caption{Riccati amplitudes for incoming and outgoing
trajectories for an NSN structure with two barriers. The arrows
define quasiparticle momentum directions. We also indicate
relevant Riccati amplitudes and distribution functions
parameterizing the Green-Keldysh function for the corresponding
trajectory.} \label{traject_b}
\end{figure*}

The above matrix Green-Keldysh functions can be conveniently parameterized by
four Riccati amplitudes $\gamma^{R,A}$, $\tilde \gamma^{R,A}$ and two
``distribution functions'' $x^K$, $\tilde x^K$ (here we follow the
notations adopted in Ref. \cite{Eschrig00}):
\begin{equation}
\hat g^K=
2
\hat N^R
\begin{pmatrix}
x^K - \gamma^R  \tilde x^K  \tilde \gamma^A &
-\gamma^R  \tilde x^K + x^K  \gamma^A \\
-\tilde \gamma^R  x^K + \tilde x^K  \tilde \gamma^A &
\tilde x^K - \tilde \gamma^R  x^K  \gamma^A \\
\end{pmatrix}
\hat N^A,
\label{gkparam}
\end{equation}
where functions $\gamma^{R,A}$ and $\tilde \gamma^{R,A}$ are Riccati amplitudes
\begin{equation}
\hat g^{R,A}=\pm
    \hat N^{R,A}
    \begin{pmatrix}
    1+\gamma^{R,A} \tilde \gamma^{R,A} & 2\gamma^{R,A} \\
    -2 \tilde \gamma^{R,A} & -1- \tilde \gamma^{R,A}  \gamma^{R,A} \\
    \end{pmatrix},
    \label{graparam}
\end{equation}
and $\hat N^{R,A}$ are the following matrices
\begin{equation}
\hat N^{R,A}=
    \begin{pmatrix}
    (1-\gamma^{R,A} \tilde \gamma^{R,A})^{-1} & 0 \\
    0 & (1-\tilde \gamma^{R,A}  \gamma^{R,A} )^{-1} \\
    \end{pmatrix},
    \label{nrparam}
\end{equation}
With the aid of the above parameterization one can identically transform
the quasiclassical equations \eqref{Eil} into the following set of
effectively decoupled equations for
Riccati amplitudes and distribution functions \cite{Eschrig00}
\begin{gather}
\begin{split}
i\bm{v}_F \nabla \gamma^{R,A}+ [\varepsilon+eV(\bm{r},t)]\gamma^{R,A}+
\gamma^{R,A}[\varepsilon-eV(\bm{r},t)]
\\
=\gamma^{R,A}\Delta^* i\sigma_2\gamma^{R,A}-\Delta i\sigma_2,
\end{split}
\label{eqgamma}
\\
\begin{split}
i\bm{v}_F \nabla \tilde\gamma^{R,A}- [\varepsilon-eV(\bm{r},t)]\tilde\gamma^{R,A}-
\tilde\gamma^{R,A}[\varepsilon+eV(\bm{r},t)]
\\
=\tilde\gamma^{R,A}\Delta i\sigma_2\tilde\gamma^{R,A}-\Delta^* i\sigma_2,
\end{split}
\label{eqtildegamma}
\\
\begin{split}
i\bm{v}_F \nabla
x^K+[\varepsilon+eV(\bm{r},t)]x^K-x^K[\varepsilon+eV(\bm{r},t)]
\\
-\gamma^{R}\Delta^* i\sigma_2 x^K-
x^K \Delta i\sigma_2 \tilde\gamma^{A}=0,
\end{split}
\label{eqx}
\\
\begin{split}
i\bm{v}_F \nabla \tilde x^K-
[\varepsilon-eV(\bm{r},t)]\tilde x^K+\tilde x^K[\varepsilon-eV(\bm{r},t)]
\\
-\tilde \gamma^{R}\Delta i\sigma_2 \tilde x^K-
\tilde x^K \Delta^* i\sigma_2 \gamma^{A}=0.
\end{split}
\label{eqtildex}
\end{gather}

Depending on the particular trajectory it is also convenient to
introduce an additional ``replica'' of both Riccati amplitudes and
distribution
functions which -- again following the notations adopted in Refs.
\cite{Eschrig00,Zhao04}) will be denoted by capital letters $\Gamma$
and $X$. These ``capital'' Riccati amplitudes and
distribution functions obey the same \eqref{eqgamma}-\eqref{eqtildex}
with the replacement $\gamma \rightarrow \Gamma$ and $x \rightarrow X$.
The distinction between different Riccati amplitudes and distribution
functions will be made explicit below.

\subsection{Boundary conditions}

Quasiclassical equations should be supplemented by
appropriate boundary conditions at metallic interfaces.
These conditions were initially formulated by Zaitsev \cite{Zaitsev}
and later generalized to the case of spin-active
(and specularly reflecting) interfaces in Ref. \cite{Millis88}.

Before specifying these conditions we would like to note that the
applicability of the above quasiclassical formalism with
appropriate boundary conditions to hybrid structures with two (or
more) barriers is, in general, a non-trivial issue \cite{GZ02,OS}
which requires a comment. Electrons scattered at different
barriers may interfere and form bound states (resonances) which in
general cannot be correctly described within the quasiclassical
formalism employing Zaitsev boundary conditions. In our geometry,
however, any relevant trajectory reaches each interface only once
whereas the probability of multiple reflections at both interfaces
is small in the parameter $\mathcal{A}_1 \mathcal{A}_2/L^4 \ll 1$.
Hence, resonances formed by multiply reflected electron waves can
be neglected, and our formalism remains adequate for the problem
in question.

In what follows we will make use of boundary conditions formulated directly
in terms of Riccati amplitudes and the distribution
functions. In the case of spin-active interfaces these
conditions are rather cumbersome \cite{Zhao04} and therefore are relegated to
Appendix.

In our three terminal geometry nonlocal conductance arises only from
trajectories that cross both interfaces. Consider first the contribution
of trajectories illustrated in Fig.~\ref{traject_b} where, for simplicity, 
we assume identical polarizations of both interfaces SN$_1$ and SN$_2$. 
In this case
the scattering matrix $\mathcal{S}$ for the first interface is defined in
Eq.~\eqref{smatrix} while the $\mathcal{S}$-matrix for the second interface
can be obtained from Eq.~\eqref{smatrix} by means of the replacement
$1 \rightarrow 2$. Accordingly, all Riccati amplitudes
have zero diagonal elements and all distribution functions have
zero off-diagonal elements.

Finally, one needs to specify the asymptotic boundary conditions
far from NS interfaces. Deep in metallic electrodes we have
\begin{gather}
\gamma_1^R=\tilde \gamma_1^R=\gamma_1^A=\tilde \gamma_1^A=0,
\\
x_1^K=\tanh\left[\dfrac{\varepsilon+eV_1}{2T}\right],\quad
\tilde x_1^K=-\tanh\left[\dfrac{\varepsilon-eV_1}{2T}\right],
\\
\gamma_2^R=\tilde \gamma_2^R=\gamma_2^A=\tilde \gamma_2^A=0,
\\
x_2^K=\tanh\left[\dfrac{\varepsilon+eV_2}{2T}\right],
\tilde x_2^K=-\tanh\left[\dfrac{\varepsilon-eV_2}{2T}\right],
\end{gather}
while in the bulk of superconducting electrode we have
\begin{gather}
\tilde \gamma_{1'}^R=-a(\varepsilon)i\sigma_2, \quad
\gamma_{1'}^A=a^*(\varepsilon)i\sigma_2,
\\
\tilde x_{1'}^K=-[1-|a(\varepsilon)|^2]\tanh\dfrac{\varepsilon}{2T},
\\
\gamma_{2'}^R=a(\varepsilon)i\sigma_2, \quad
\tilde\gamma_{2'}^A=-a^*(\varepsilon)i\sigma_2,
\\
x_{2'}^K=[1-|a(\varepsilon)|^2]\tanh\dfrac{\varepsilon}{2T},
\end{gather}
where $a( \varepsilon ) = - (\varepsilon - \sqrt{ \varepsilon^2 -
\Delta^2})/\Delta$.

\subsection{Green functions and non-local conductance}

With the aid of the above equations and boundary conditions it is
straightforward to evaluate the quasiclassical Green-Keldysh functions
for our three-terminal device along whole trajectory. For
instance, from the boundary conditions on the second interface we find
\begin{equation}
\Gamma_{2'}^R=\sqrt{R_{2\uparrow}R_{2\downarrow}}
\begin{pmatrix}
e^{i\theta_2} & 0 \\
0 & e^{-i\theta_2} \\
\end{pmatrix}
i\sigma_2 a(\varepsilon).
\label{GGG}
\end{equation}
Then integrating equation \eqref{eqgamma} along trajectory
connecting both interfaces and using Eq. (\ref{GGG}) as the
initial condition we arrive at the Riccati amplitude at the first
interface:
$$
\gamma_{1'}^R=
\dfrac{\Gamma_{2'}^R+(\varepsilon\Gamma_{2'}^R+\Delta i\sigma_2)Q}{
1-(\varepsilon-\Delta \Gamma_{2'}^R i\sigma_2)Q},
$$
\begin{equation}
Q=\dfrac{\tanh\left[i\Omega L/v_F\right]}{
\Omega}, \quad
\Omega = \sqrt{\varepsilon^2-\Delta^2}.
\end{equation}
Similarly, integrating Eq. \eqref{eqx}, one finds
\begin{equation}
x_{1'}^K=X_{2'}^K\dfrac{1-\gamma_{1'}^R \tilde\gamma_{1'}^A}{
1-\Gamma_{2'}^R\tilde\Gamma_{2'}^A}.
\end{equation}
We also note that the relation $(\gamma^{R,A})^+ = \tilde \Gamma^{A,R}$
makes it unnecessary (while redundant) to separately
calculate the amplitudes $\tilde\gamma_{1'}^A$ and $\tilde\Gamma_{2'}^A$.

Finally we arrive at the following general expression for the
Green-Keldysh function $g^K_{1}$ at SN$_1$ interface \cite{KZ06}
\begin{equation}
g_1^K=g_{11}^K(V_1)+\delta g_{12}^K(V_1)+\delta g_{1}^K(V_2),
\label{gN1}
\end{equation}
where the three different terms in the right-hand side correspond
respectively to local contribution and two non-local corrections
depending on $V_1$ and $V_2$. Since here we are only interested in
the non-local conductance of our
 structure, it is sufficient to keep only the part of the Green-Keldysh
function $\delta g^K_1$ which depends on the voltage
 $V_2$ across the second interface.
Combining the above results with the boundary conditions for
the distribution functions $X_{2'}^K$ and
$X_1^K$ at both interfaces respectively one immediately recovers
$\delta g^K_1$. E.g., for the outgoing momenta directions ($1^{out}$)
one gets
\begin{multline}
 \delta g^{K(1)}_1=\delta X_1^K =
2 (1-\tanh^2iL\Omega/v_F) \\
\times \begin{pmatrix}
1/P(\varepsilon,z_1,z_2) & 0 \\
0 & 1/P(-\varepsilon,z_1,z_2) \\
\end{pmatrix}
\\\times \left\{
\begin{pmatrix}
D_{1\uparrow} D_{2\uparrow} & 0 \\
0 &  D_{1\downarrow} D_{2\downarrow}\\
\end{pmatrix}
\tanh\dfrac{\varepsilon+eV_2}{2T}
\right.
\\+
\left.
|a|^2
\begin{pmatrix}
D_{1\uparrow} D_{2\downarrow}R_{2\uparrow} & 0 \\
0 &  D_{1\downarrow} D_{2\uparrow}R_{2\downarrow}\\
\end{pmatrix}
\tanh\dfrac{\varepsilon-eV_2}{2T}
\right\},
\label{gkb}
\end{multline}
where $\delta X_1^K$ is the $V_2$-dependent part of the $X_1^K$
distribution function, $z_i=\sqrt{ R_{i\uparrow} R_{i\downarrow} }
\exp( i \theta_i)$, $i=1,2$ $R_{i\uparrow (\downarrow )}$ are the
spin-sensitive reflection coefficients, $\theta_i$ are the
spin-mixing angles and
\begin{multline}
P(\varepsilon,z_1,z_2)=
\Biggl|
1-z_1 z_2 a^2
-\dfrac{\tanh iL\Omega/v_F}{\Omega}\\\times
\left[
\varepsilon
(1+z_1 z_2 a^2)
+a\Delta (z_1 + z_2)
\right]
\Biggr|^2.
\end{multline}

Reversing momenta directions for all trajectories in
Fig.~\ref{traject_b} yields the second set of trajectories which
also contributes to the non-local conductance. The corresponding
expression for the $V_2$-dependent part of the Green-Keldysh
function $\delta g^{K(2)}_1$ at the normal metal side of the first
interface is derived analogously to Eq. (\ref{gkb}). Again for the
outgoing momenta directions we obtain
\begin{multline}
 \delta g^{K(2)}_1=
2 (1-\tanh^2iL\Omega/v_F) \\\times
\begin{pmatrix}
1/P(\varepsilon,z_1,z_2) & 0 \\
0 & 1/P(-\varepsilon,z_1,z_2) \\
\end{pmatrix}
\\\times \left\{ |a|^4
\begin{pmatrix}
D_{1\uparrow} R_{1\downarrow}D_{2\uparrow}R_{2\downarrow} & 0 \\
0 &  D_{1\downarrow} R_{1\uparrow} D_{2\downarrow} R_{2\uparrow}\\
\end{pmatrix}
\tanh\dfrac{\varepsilon+eV_2}{2T}
\right.
\\+
\left.
|a|^2
\begin{pmatrix}
D_{1\uparrow} R_{1\downarrow}D_{2\downarrow} & 0 \\
0 &  D_{1\downarrow} R_{1\uparrow} D_{2\uparrow}\\
\end{pmatrix}
\tanh\dfrac{\varepsilon-eV_2}{2T}
\right\}.
\label{gkc}
\end{multline}
Finally we note that in the expression for function $g^K$ there
also exist terms corresponding to incoming momenta directions
(again on the normal metal side of the first interface). However,
those terms do not depend on the voltage $V_2$ and, hence, do not
contribute to the non-local conductance at all. Thus, Eqs.
\eqref{gkb} and \eqref{gkc} already contain complete information
which allows to evaluate the non-local current across the first
interface. This task will be accomplished below.

\section{Nonlocal conductance}
\subsection{General results}
With the aid of general expressions for the Green-Keldysh
functions obtained here it is possible to demonstrate that the
total currents across the first and the second interface NS
interfaces $I_1$ and $I_2$ have the form \cite{KZ06}
\begin{gather}
I_1= I_{11}(V_1)+I_{12}(V_2),
\\
I_2= I_{21}(V_1)+I_{22}(V_2).
\end{gather}
Here we are interested in the non-local part of the current
$I_{12}(V_2)$ across the first interface. Substituting \eqref{gkb}
and \eqref{gkc} into \eqref{current} we obtain
\begin{multline}
I_{12}(V_2)=- \dfrac{G_0}{4e} \int d \varepsilon
\left[\tanh\dfrac{\varepsilon+eV_2}{2T} -
\tanh\dfrac{\varepsilon}{2T} \right] \\\times
(1-\tanh^2iL\Omega/v_F) \\\times \left( \dfrac{ D_{1\uparrow}
D_{2\uparrow}[1+ |a|^4
R_{1\downarrow}R_{2\downarrow}]-|a|^2(D_{1\downarrow}D_{2\uparrow}
[R_{2\downarrow}+ R_{1\uparrow} ] }{P(\varepsilon,z_1,z_2)}
\right.
\\+
\left.
\dfrac{
D_{1\downarrow} D_{2\downarrow}[1
+|a|^4 R_{1\uparrow} R_{2\uparrow}]-|a|^2(D_{1\uparrow}
D_{2\downarrow}[R_{2\uparrow}
+R_{1\downarrow}]
}{P(-\varepsilon,z_1,z_2)}
\right)
\label{I12}
\end{multline}
where
\begin{equation}
G_{N_{12}}=G_0 \dfrac{D_{1\uparrow} D_{2\uparrow} +
D_{1\downarrow} D_{2\downarrow}}{2}
\end{equation}
is the non-local conductance of our device in the normal state,
$$
G_0=\frac{8\gamma_1 \gamma_2
\mathcal{N}_1\mathcal{N}_2}{R_qp_F^2L^2},
$$
$p_F\gamma _{1(2)}$ is normal to the first (second)
interface component of the Fermi momentum for electrons
propagating straight between the interfaces, 
$\mathcal{N}_{1,2}=p_F^2\mathcal{A}_{1,2}/4\pi$ define the number of conducting
channels of the corresponding interface, $R_q=2\pi/e^2$ is the
quantum resistance unit.

Eq. \eqref{I12} is the central result of our paper. It fully
determines the non-local spin-dependent current in our
three-terminal ballistic NSN structure at arbitrary interface
transmissions, voltages and temperature. It is also convenient 
to introduce the non-local differential conductance 
$G_{12}(V_2)=-\partial I_{12}/\partial V_2$.
In the limit $T, V_{1,2}\ll \Delta$ only subgap quasiparticles
contribute to the current and the differential conductance becomes
voltage-independent. We have $I_{12}=-G_{12}V_{2}$, where
\begin{multline}
G_{12}= G_0 (1-\tanh^2L\Delta/v_F) \\\times
\dfrac{(D_{1\uparrow}-D_{1\downarrow})(D_{2\uparrow}-D_{2\downarrow})+
D_{1\uparrow}D_{1\downarrow}D_{2\uparrow}D_{2\downarrow}
}{P(0,z_1,z_2)} \label{G12zeroV}.
\end{multline}
For spin-isotropic interfaces Eqs. \eqref{G12zeroV} and
\eqref{I12} yield the results obtained previously in \cite{KZ06}.
In the lowest (first order) order in the transmissions of both
interfaces Eq. \eqref{G12zeroV} reduces to the result by Falci
{\it et al.} \cite{FFH} provided we assume that the ratio 
$D_{1(2)\uparrow}/D_{1(2)\downarrow}$ coincides with that of 
spin-up and spin-down densities of states in the first (second) 
ferromagnetic electrode.
We also note that, provided at least one of the interfaces is
spin-isotropic, the conductance \eqref{G12zeroV} is proportional
to the product of all four transmissions
$D_{1\uparrow}D_{1\downarrow}D_{2\uparrow}D_{2\downarrow}$, i.e.
it differs from zero only due to processes involving scattering
with both spin projections in both normal electrodes.

As in the case of spin-isotropic interfaces the value $G_{12}$
\eqref{G12zeroV} gets strongly suppressed with increasing $L$, and
this dependence on $L$ is in general non-exponential reducing to
$G_{12} \propto \exp (-2L\Delta /v_F)$ either in the limit of
small transmissions or large  $L \gg v_F/\Delta$.

In the spin-degenerate case for a given $L$ the non-local
conductance reaches its maximum reflectionless barriers
$D_{1,2}=1$. Interestingly, in this case for small $L \ll
v_F/\Delta$ the conductance $G_{12}$ identically coincides with
its normal state value $G_{N_{12}}$ at any temperature and
voltage. For $L\to 0$ there is ``no space'' for CAR to develop on
these trajectories and, hence, CAR contribution to $G_{12}$
vanishes, whereas direct transfer of electrons between N$_1$ and
N$_2$ remains unaffected by superconductivity in this limit.

The situation changes provided at least one of the transmissions
is smaller than one. In this case scattering at SN interfaces
mixes up trajectories connecting N$_1$ and N$_2$ terminals with
ones going deep into and coming from the superconductor. As a
result, CAR contribution to $G_{12}$ does not vanish even in the
limit $L \to 0$ and $G_{12}$ turns out to be smaller than
$G_{N_{12}}$.

\subsection{Polarized interfaces}
\begin{figure}
\centerline{
\includegraphics[width=75mm]{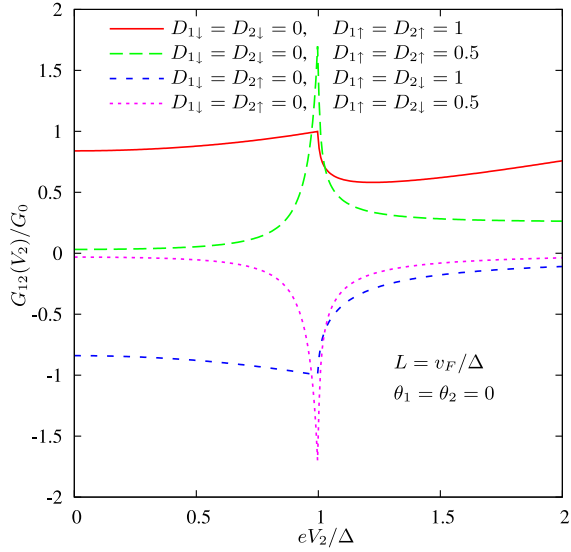}
} \caption{Zero temperature differential non-local conductance as
a function of voltage for a half-metal/superconductor/half-metal
structure. Different curves correspond to different interface
transmissions and different orientations of the half-metal
magnetizations.} \label{current-hsh}
\end{figure}

Let us now turn to the limit of highly polarized interfaces which
is accounted for by either spin-up or spin-down transmission of
each interface going to zero. In this limit our model describes an
HSH structure, where H stands for fully spin-polarized half-metallic 
electrodes. In
the case of parallel magnetization of both half-metals
($D_{1\uparrow}=D_1$, $D_{1\downarrow}=0$, $D_{2\uparrow}=D_2$,
and $D_{2\downarrow}=0$) we obtain
\begin{multline}
I_{12}^{\uparrow\uparrow}(V_2)=- \dfrac{G_0}{4e} \int d
\varepsilon \left[\tanh\dfrac{\varepsilon+eV_2}{2T} -
\tanh\dfrac{\varepsilon}{2T} \right] \\\times
(1-\tanh^2iL\Omega/v_F) \dfrac{ D_{1} D_{2} (1+|a|^4 )
}{P(\varepsilon,z_1,z_2)}, \label{hsh1}
\end{multline}
while in the case of antiparallel magnetization
($D_{1\uparrow}=D_1$, $D_{1\downarrow}=0$, $D_{2\uparrow}=0$, and
$D_{2\downarrow}=D_2$) we find
\begin{multline}
I_{12}^{\uparrow\downarrow}(V_2)=+ \dfrac{G_0}{4e} \int d
\varepsilon \left[\tanh\dfrac{\varepsilon+eV_2}{2T} -
\tanh\dfrac{\varepsilon}{2T} \right] \\\times
(1-\tanh^2iL\Omega/v_F) \dfrac{ D_{1} D_{2} 2 |a|^2
}{P(-\varepsilon,z_1,z_2)}, \label{hsh2}
\end{multline}
i.e. the non-local conductances \eqref{hsh1} and \eqref{hsh2} have
opposite signs. Some typical curves for the differential non-local
conductance are presented in Fig.~\ref{current-hsh} at
sufficiently high interface transmissions and zero spin mixing angles. With decreasing
interface transmissions low voltage non-local conductance
diminishes and sharp peaks at voltages $eV_2=\pm\Delta$ appear.

\begin{figure}
\centerline{
\includegraphics[width=75mm]{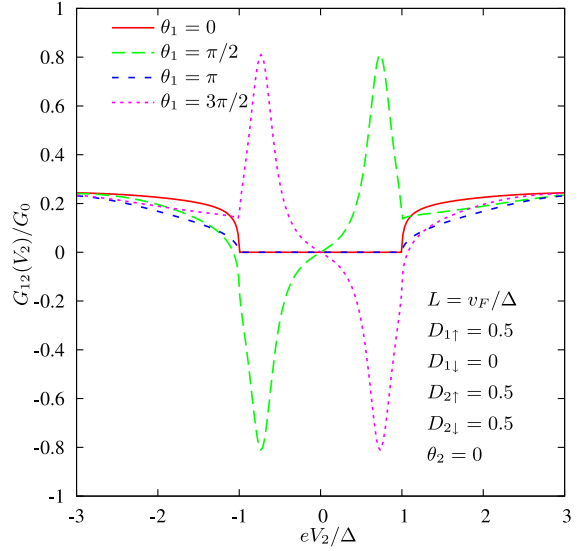}
} \caption{Zero temperature differential non-local conductance as
a function of voltage for a half-metal/superconductor/metal
structure. Different curves correspond to different values of the
spin-mixing angle $\theta_1$.} \label{current-hsn}
\end{figure}

We will now turn to  asymmetric HSN heterostructures with one
half-metallic and one spin-isotropic electrode. In this case the
non-local current does not depend on the magnetization  direction of the
half-metal. Below we will distinguish NSH and HSN configurations.
In the case of metal/superconductor/half-metal (NSH) structures
($D_{1\uparrow}=D_{1\downarrow}=D_1$, $D_{2\uparrow}=D_2$,
$D_{2\downarrow}=0$) we obtain
\begin{multline}
I_{12}(V_2)=- \dfrac{G_0}{4e} \int d \varepsilon
\left[\tanh\dfrac{\varepsilon+eV_2}{2T} -
\tanh\dfrac{\varepsilon}{2T} \right] \\\times
(1-\tanh^2iL\Omega/v_F) \dfrac{ D_{1} D_{2} (1-|a|^2 ) (1-R_1|a|^2
) }{P(\varepsilon,z_1,z_2)}.
\label{nsh}
\end{multline}

\begin{figure}
\centerline{
\includegraphics[width=75mm]{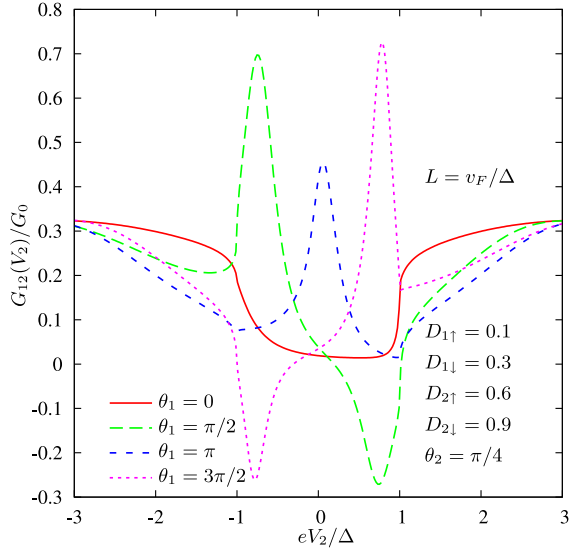}
} \caption{Zero temperature differential non-local conductance as
a function of voltage for a general case of an NSN structure with
non-zero transmissions $D_{1\uparrow}$, $D_{1\downarrow}$,
$D_{2\uparrow}$ and $D_{2\downarrow}$ and non-zero spin-mixing
angles $\theta_1$ and $\theta_2$.} \label{current-gen}
\end{figure}

We note that in this case the spectral current is identically zero
at all subgap energies $|\varepsilon| < \Delta$. The sign of the
non-local current and differential conductance remain the same as
in the normal state. In the reversed situation of a
half-metal/superconductor/metal (HSN) structure
($D_{1\uparrow}=D_1$, $D_{1\downarrow}=0$,
$D_{2\uparrow}=D_{2\downarrow}=D_2$) our general results yield
\begin{multline}
I_{12}(V_2)=- \dfrac{G_0}{4e} \int d \varepsilon
\left[\tanh\dfrac{\varepsilon+eV_2}{2T} -
\tanh\dfrac{\varepsilon}{2T} \right] \\\times D_{1}
D_{2}(1-\tanh^2iL\Omega/v_F) \\
\times \left( \dfrac{1+|a|^4 R_{2}}{P(\varepsilon,z_1,z_2)} -
\dfrac{|a|^2(1+R_{2})}{P(-\varepsilon,z_1,z_2)} \right).
\label{hsn}
\end{multline}
In contrast to Eq. (\ref{nsh}), here the spectral current does not
vanish at subgap energies provided the spin-mixing angle
$\theta_1$ differs from zero. The sign of the non-local
conductance in this geometry depends on both applied voltage $V_2$
and spin-mixing angle $\theta_1$. The zero temperature non-local
conductance evaluated with the aid of Eq. \eqref{hsn} at
sufficiently high interface transmissions and different
spin-mixing angles is depicted in Fig. \ref{current-hsn}.

In general, the voltage dependence of the non-local current is
very sensitive to the interface transmissions and -- in particular
-- to values of spin-mixing angles $\theta_{1,2}$.  Typical curves
illustrating the voltage dependence of the differential non-local
conductance are shown in Fig.~\ref{current-gen} at different
spin-mixing angles. Positions of the peaks correspond to energies
of the quasibound states in NSN structures.

\section{Conclusions}

In this paper we have developed a non-perturbative theory of
spin-resolved non-local electron transport in ballistic NSN
three-terminal structures with spin-active interfaces. Our theory
applies at arbitrary interface transmissions and allows to fully
describe a non-trivial interplay between spin-sensitive normal
scattering, local and non-local Andreev reflection at SN
interfaces. We have evaluated the non-local conductance of our NSN
device at arbitrary voltages and temperatures and observed a
number of interesting properties of such structures with
spin-active interfaces. Our results can be applied to various NSN
hybrid structures, including systems with ferromagnetic and
half-metallic electrodes, and can be directly tested in future
experiments.

\appendix

\section{Boundary conditions for spin-active interfaces}

\begin{figure}
\centerline{\includegraphics[width=75mm]{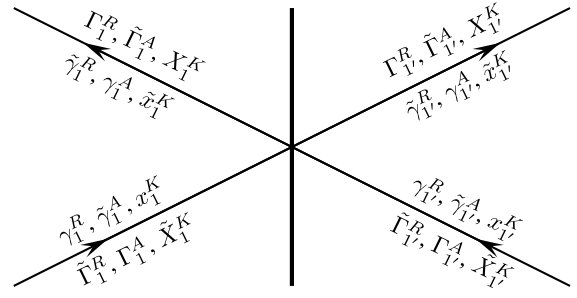}}
\caption{Riccati amplitudes for incoming and outgoing trajectories from the both
sides of the interface. }
\label{sis}
\end{figure}

Let us explicitly specify the relations between Riccati amplitudes
and distribution functions for incoming and outgoing trajectories
at the first interface, see Fig.~\ref{sis}. The boundary
conditions for $\Gamma_1^R$, $\tilde\Gamma_1^A$ and $X_1^K$ can be
written in the form \cite{Zhao04}
\begin{gather}
\Gamma_1^R= r_{1l}^R\gamma_1^R \underline{S}_{11}^+ +
t_{1l}^R\gamma_{1'}^R {\underline{S}}_{11'}^+,
\\
\tilde\Gamma_1^A= \underline{S}_{11} \tilde\gamma_1^A \tilde r_{1r}^A  +
{\underline{S}}_{11'} \tilde\gamma_{1'}^A \tilde t_{1r}^A ,
\\
X_1^K=r_{1l}^R x_1^K \tilde r_{1r}^A +
t_{1l}^R x_{1'}^K \tilde t_{1r}^A -
a_{1l}^R \tilde x_{1'}^K \tilde a_{1r}^A.
\end{gather}
Here the transmission ($t$), reflection  ($r$), and
branch-conversion ($a$) amplitudes are defined as follows:
\begin{gather}
r_{1l}^R=+[(\beta_{1'1}^R)^{-1}S_{11}^+ - (\beta_{1'1'}^R)^{-1}S_{11'}^+]^{-1}
(\beta_{1'1}^R)^{-1},
\\
t_{1l}^R=-[(\beta_{1'1}^R)^{-1}S_{11}^+ - (\beta_{1'1'}^R)^{-1}S_{11'}^+]^{-1}
(\beta_{1'1'}^R)^{-1},
\\
\tilde r_{1r}^A=+(\beta_{1'1}^A)^{-1}
[S_{11}(\beta_{1'1}^A)^{-1} - S_{11'}(\beta_{1'1'}^A)^{-1}]^{-1},
\\
\tilde t_{1r}^A=-(\beta_{1'1'}^A)^{-1}
[S_{11}(\beta_{1'1}^A)^{-1} - S_{11'}(\beta_{1'1'}^A)^{-1}]^{-1},
\\
a_{1l}^R=(\Gamma_1^R \underline{S}_{11} - S_{11}\gamma_1^R)(\tilde
\beta_{11'}^R)^{-1},
\\
\tilde a_{1r}^A=(\tilde \beta_{11'}^A)^{-1}
(\underline{S}_{11}^+ \tilde \Gamma_1^A  - \tilde \gamma_1^A S_{11}^+),
\end{gather}
where
\begin{gather}
\beta_{ij}^R=S_{ij}^+ - \gamma_j^R \underline{S}_{ij}^+ \tilde\gamma_i^R,\
\tilde\beta_{ij}^R=\underline{S}_{ji} - \tilde \gamma_j^R S_{ji}\gamma_i^R,
\\
\beta_{ij}^A=S_{ij} - \gamma_i^A \underline{S}_{ij} \tilde\gamma_j^A,\
\tilde\beta_{ij}^A=\underline{S}_{ji}^+ - \tilde \gamma_i^A S_{ji}^+\gamma_j^A.
\end{gather}
Similarly, the boundary conditions for $\tilde\Gamma_1^R$,
$\Gamma_1^A$, and $\tilde X_1^K$ are:
\begin{gather}
\tilde\Gamma_1^R= \tilde r_{1l}^R\tilde\gamma_1^R S_{11} +
\tilde t_{1l}^R\tilde\gamma_{1'}^R S_{1'1},
\\
\Gamma_1^A= S_{11}^+ \gamma_1^A r_{1r}^A  +
S_{1'1}^+ \gamma_{1'}^A t_{1r}^A ,
\\
\tilde X_1^K=\tilde r_{1l}^R \tilde x_1^K r_{1r}^A +
\tilde t_{1l}^R \tilde x_{1'}^K t_{1r}^A -
\tilde a_{1l}^R x_{1'}^K a_{1r}^A,
\end{gather}
where
\begin{gather}
\tilde r_{1l}^R=+[(\tilde \beta_{1'1}^R)^{-1}\underline{S}_{11} -
(\tilde \beta_{1'1'}^R)^{-1}\underline{S}_{1'1}]^{-1}
(\tilde\beta_{1'1}^R)^{-1},
\\
t_{1l}^R=-[(\tilde \beta_{1'1}^R)^{-1}\underline{S}_{11} -
(\tilde \beta_{1'1'}^R)^{-1}\underline{S}_{1'1}]^{-1}
(\tilde\beta_{1'1'}^R)^{-1},
\\
r_{1r}^A=+(\tilde\beta_{1'1}^A)^{-1}
[\underline{S}_{11}^+(\tilde\beta_{1'1}^A)^{-1} -
\underline{S}_{1'1}^+(\tilde\beta_{1'1'}^A)^{-1}]^{-1},
\\
\tilde t_{1r}^A=-(\tilde\beta_{1'1'}^A)^{-1}
[\underline{S}_{11}^+(\tilde\beta_{1'1}^A)^{-1} -
\underline{S}_{1'1}^+(\tilde\beta_{1'1'}^A)^{-1}]^{-1},
\\
a_{1l}^R=(\Gamma_1^R \underline{S}_{11} - S_{11}\gamma_1^R)(\tilde
\beta_{11'}^R)^{-1},
\\
\tilde a_{1r}^A=(\tilde \beta_{11'}^A)^{-1} (\underline{S}_{11}^+
\tilde \Gamma_1^A  - \tilde \gamma_1^A S_{11}^+).
\end{gather}
Boundary conditions for $\Gamma_{1'}^{R,A}$, $\tilde \Gamma_{1'}^{R,A}$, $X^K_{1'}$ 
and $\tilde X^K_{1'}$ can be obtained from the above equations by means of the
replacement $1 \leftrightarrow 1'$.

Matrices $S_{11}$, $S_{11'}$, $S_{1'1}$, and $S_{1'1'}$ are the
components of the $\mathcal{S}$-matrix describing electron
scattering at the first interface:
\begin{equation}
\mathcal{S}=
\begin{pmatrix}
S_{11} & S_{11'}\\
S_{1'1} & S_{1'1'}\\
\end{pmatrix}, \quad
\mathcal{S}\mathcal{S}^+=1
\end{equation}
For specularly reflecting interfaces with inversion symmetry
elements of the $\mathcal{S}$-matrix have the following form
\begin{gather}
S_{11}=S_{1'1'}=
\begin{pmatrix}
\sqrt{R_{1\uparrow}}e^{i\theta_1/2} & 0 \\
0 & \sqrt{R_{1\downarrow}}e^{-i\theta_1/2} \\
\end{pmatrix},
\\
S_{11'}=S_{11'}=i
\begin{pmatrix}
\sqrt{D_{1\uparrow}}e^{i\theta_1/2} & 0 \\
0 & \sqrt{D_{1\downarrow}}e^{-i\theta_1/2} \\
\end{pmatrix},
\label{smatrix}
\end{gather}
where $D_{1\uparrow}$ and $D_{1\downarrow}$ are interface
transmission for spin-up and spin-down electron. The spin-mixing
angle $\theta_1$ accounts for the difference between scattering
phases for processes with opposite spin directions. The condition
$\underline{S}_{ij}={S}_{ij}$ holds provided there exists
reflection symmetry in the plane perpendicular to the interface.

\end{document}